\documentclass[prd,preprintnumbers,nofootinbib,aps,twocolumn]{revtex4-1}
\usepackage{epsfig,amsmath,amssymb}
\usepackage{mathrsfs}
\usepackage[normalem]{ulem}
\usepackage[usenames,dvipsnames]{color}
\usepackage[pagebackref=false, colorlinks=false]{hyperref}

\begin{document}
\title{Inverse Area Corrections to Black Hole Entropy Area Formula in classical F(R) Gravity and Loop Quantum Gravity, and Gravitational Wave Observations}

\author{Rohit Das{\email{rodas21965@gmail.com}}, Parthasarathi Majumdar\email{bhpartha@gmail.com}, Debadrita Mukherjee{\email{mukherjeedebadrita531@gmail.com}} }

\affiliation{School of Physical Sciences, Indian Association for the Cultivation of Science, Kolkata 700032, India.} 

\begin{abstract}

We consider corrections to the Bekenstein-Hawking Area Formula (BHAF) for black hole entropy, which have {\it inverse powers} of the horizon area for very large horizon areas, for classical spherically symmetric black hole solutions of $F(R)$ modified gravity theory, using the Wald formula for the entropy function with modifications suggested by Jacobson, Kang and Myers. Requiring that the coefficient of such corrections be {\it absolutely} consistent with gravitational wave observational results validating Hawking's Area Theorem for binary black hole coalescences, implies constraints on parameters of $F(R)$ gravity. For the sake of comparison, we present a computation of inverse area corrections for quantum black holes in {\it quantum general relativity}, using Wheeler's It from Bit approach modified by some tenets of Loop Quantum Gravity. 
 \end{abstract}

\maketitle

\section{Introduction}

Consistency of the logrithmic corrections to the Bekenstein-Hawking Area Formula (BHAF) \cite{bek73}-\cite{haw75} for black hole entropy, with gravitational wave (GW) observational results \cite{teu20}-\cite{abac25} validating Hawking's Area Theorem (HAT) \cite{haw71} for astrophysical binary black hole coalescences (BBHC), have been considered in \cite{pm2024}, \cite{pm2025} and \cite{pm2025-2} by one of us (PM). In ref. \cite{pm2025-2} an attempt is made in combining the formulas derived through Loop Quantum Gravity (LQG) \cite{km98}-\cite{abhi-pm14} and also the perturbative one loop entanglement entropy formula based on Euclidean Quantum Gravity \cite{cal1994}-\cite{sol2020}. The absolute consistency requirement of the {\it overall} logarithmic corrections, with GW observations on BBHC, is seen to lead to implications on the number of species of matter field fluctuations, with spin-parity profiles within and beyond the spectrum of fields in the Standard Strong-Electroweak Theory of elementary particle physics. 

As argued already in ref. \cite{pm2025-2}, logarithmic corrections to the BHAF ensue from quantum formulations of general relativity and interacting matter fields. On the other hand, if we restrict our attention to classical modifications of general relativity, corrections to the BHAF are most likely to occur as a power series in {\it inverse} horizon area, for astrophysical black holes of macroscopic horizon areas. The coefficients of such inverse powers of area correction will of course depend on parameters of the modified gravity theory under consideration. The notion of absolute consistency with the GW observations on BBHC validating the HAT, proposed in ref.s \cite{pm2024}-\cite{pm2025-2} incorporating the Generalized Second Law of Thermodynamics \cite{bek73} (GSLT), is likely therefore to restrict parameters of the modified gravity theory being considered. 

The primary methodology for computing the entropy for black hole solutions of modified gravity theories with diffeomorphism invariance, is the application of the Wald entropy function \cite{rmw1992}, \cite{iw1994} formulated as a N\"other charge. While the original formulation of ref.s \cite{rmw1992}, \cite{iw1994} involve the computation of the N\"other charge as an integral over the bifurcation of the Killing horizon of a stationary black hole, Jacobson et. al. \cite{jkm1994} demonstrate that Wald's entropy function may be generalized to an integral of the conserved N\"other current over {\it any} two dimensional cross-section of the horizon, including by a spatial foliation of the horizon. It is this modification that we find particularly useful to estimate the inverse powers of area correction to the area formula for entropy of stationary spherical black holes of modified gravity theories. 

We may mention that there is earlier literature on the use of the Wald-Jacobson-Kang-Myers formalism to calculate the entropy of black holes in $F(R)$ gravity (without and with matter) (see, for instance, ref.s \cite{bre2004}-\cite{mod2011}), which overlap with this work. Working within an {\it arbitrary} $F(R)$ gravity framework without matter, we rederive the main formula for the entropy of vacuum static black hole solutions of the theory, first derived seventeen years ago in ref. \cite{bru2009} and reviewed in ref. \cite{far2010}\footnote{Our oversight, in an earlier version of this paper, in not recognizing the fact that ours is a {\it rederivation} of the older result of ref.s \cite{bru2009}-\cite{far2010}, is regretted.}. Our work is restricted to four dimensional spacetime and spherical black holes of the vacuum theory, and follows the generalization offered in ref. \cite{jkm1994} of the integration of the N\"other current over a codimension 2 spherical foliation of the full spacetime, to yield the entropy function.  

For astrophysically large black holes, we obtain, from the main formula for black hole entropy, an expression which closely resembles what may be construed as an analogue of the usual BHAF of general relativity. This expression is the first term in a power series in inverse horizon area, showing the emergence of a whole slew of inverse powers (in area) {\it corrections} to the BHAF for $F(R)$ gravity. The identification of the first derivative of $F(R)$ at vanishing argument thus stands directly related to the `Planck Area' $A_F$ of $F(R)$ gravity which in turn defines the effective Newtonian coupling {\it constant}. Although implicit in ref.s \cite{bru2009}-\cite{noj2011}, we believe this is the first instance in which such a result has been articulated explicitly, so that one modification to general relativity that $F(R)$ gravity entails is delineated in terms of the calculated entropy of spherical black holes, at least for astrophysically large configurations, as the corresponding BHAF + corrections. In the sequel, we shall focus on possible corrections to the BHAF, which are inverse powers of the BHAF value. Observe that we are {\it not} considering any truncation of the {\it theory} in powers of the Ricci scalar, but emphasize that the corrections emerge from the full $F(R)$ theory for a class of $F$s. One reason for avoiding truncating the theory, is our suspicion that such truncation may have an issue with the initial value formulation of the theory itself. Since we do not address such issues here, we prefer not to truncate the primary action of $F(R)$ gravity. 

Thus, in this paper, the inverse powers of area corrections to the BHAF are shown to emerge for spherical black hole solutions of $F(R)$ gravity models, for all those functions $F(R)$ which are regular for vanishingly small argument. The latter restriction is necessiated by the phenomenological requirement of large macroscopic horizon sizes, imposed by GW observations from BBHCs, which is our focus in this paper. For functions $F(R)$ which are {\it not regular} at vanishing argument, the issue of a suitable Newtonian limit lurks in the background. We do not address such issues here. E.g., one could think of $F(R) = a_1 R + a_2 \sqrt{R}$, which, even though non-divergent for real vanishing argument, has a branch cut along the real axis for complex arguments. We preclude such `non-analytic' functions $F$ from our consideration, because of ambiguities that might ensue if we look for a Newtonian gravity limit of this theory. This of course is not to claim that such choices of $F$ actually lead to any inherent pathology.     

We now turn to using results of extensive expert analyses \cite{teu20}-\cite{abac25} of LVK GW data on BBHC, which clearly validate the HAT to a very high accuracy of $5\sigma$ resulting from GW signals, especially of the postmerger phase, with SNR exceeding 80. From the HAT, combined with the GSL, we deduce an inequality on the inverse power corrections to the BHAF in terms of the observed area increase in the BBHC. Given that the inverse power corrections are tiny, compared to the BHAF for entropy, we propose the absolute consistency criterion as an outcome of the HAT validation, which may have implications for the computed corrections, and hence, on {\it some of the parameters} of $F(R)$. Albeit indirect, this linking of LVK GW data to determining parameters of $F(R)$ gravity, via inverse powers of area corrections to the BHAF for black hole entropy, has not appeared in the earlier literature and is novel in our opinion. Given the very high accuracy of the validation of the HAT, from direct analyses of GW data from BBHCs, constraints on parameters of modified gravity theories are intuitively expected. 

An objection to this conclusion may be raised on the ground that the data analyses, e.g., in \cite{abac25} on GW250114, uses key aspects of general relavity, like linear perturbations around a Kerr black hole spacetime, and not an $F(R)$ modified gravity framework. One may perhaps speculate that if the latter analysis could be performed, the constraints on the parameters of $F(R)$ gravity we have found in this paper, may ease. However, there is strictly speaking no basis to such a general remark, since analyses of GW data on the basis $F(R)$ modified gravity (for general $F$), have not yet been performed, to the best of our knowledge. Even if it were possible to perform such an analysis, it is not clear at all if the validation of a HAT-like result (or otherwise) with observational data will ever reach the $5 \sigma$ level of accuracy which has already been achieved in the GR based approach. It is hard thus to sustain the speculation without a factual basis.   

Further, the calculations of ref. \cite{bru2009}, and others, reviewed in ref. \cite{far2010}, yield, as mentioned above, to a formula for black hole entropy with close resemblances with BHAF together with corrections thereof, for large black holes. The emergence of a BHAF shows a degree of consistency with a HAT-like theorem for $F(R)$ modified gravity theories, for a class of $F$s which are regular at the origin of their argument, at least for astrophysically large black holes. Admittedly, there may be some choices of $F(R)$ for which a HAT-like result may not ensue. But are these choices free of other pathologies, like ghost freedom, which have not been carefully examined yet ? We do not know the answer to this question. Also, there exist textbook proofs of HAT based entirely on properties of null geodesics and null eneargy conditions where the Einstein equation is never explicitly used. So the domain of applicability of the HAT for BBHC events may be wider than simply the set of explicitly GR-based theoretical approaches. 

In earlier work by one of us (PM) \cite{pm2025-2}, the issue of absolute consistency of leading order logarithmic corrections to the BHAF in quantum GR has been explored. The issue of {\it subleading} corrections inverse in horizon area for large horizon areas may be raised. We understand this question may be interesting for some of the QGR proposals examined in ref. \cite{pm2025-2}. 

In this paper, the penultimate section is an excursion, outside the main purview of the paper, into emergence of an inverse area {\it subleading} correction, in {\it quantum general relativity} (QGR). Since such a correction is easy to extract in an extension of the approach of ref. \cite{pm2024}, we present it here for comparison with the classical approach to inverse area corrections to BHAF. We consider here an intuitive approach initiated by J. A. Wheeler in 1991, known as the `It from Bit' approach to compute quantum black hole entropy from a counting of states of a horizon-enveloping two dimensional lattice, if the lattice plaquettes are populated by binary bits, like spin $1/2$ variables. We find this approach fits in nicely with the LQG approach discussed in ref.s \cite{pm2024}-\cite{pm2025-2}. This LQG modification is perhaps the easiest approach to derive the LQG result for the logarithmic corrections. Here, we reexamine this approach and seek the subleading inverse area correction computed from this LQG modification of the It from Bit proposal. We then consider the effect of the absolute consistency criterion vis-a-vis GW observations, on the coefficient of the computed result, and establish such a consistency.
   
The paper is organized as follows : we briefly review the basic tenets of ref. \cite{jkm1994} in section 2. We restrict our attention, in section 3, to spherically symmetric, static black hole solutions \cite{seb2011} of $F(R)$ gravity \cite{sot2010}, and as such, the integration cross sections are 2-spheres ${\cal S}$. For large areas of these cross-sections, corresponding to astrophysical black holes of the type featuring in recently observed gravitational waves emitted in BBHCs, the Wald-Jacobson-Kang-Myers Entropy function admits a Taylor expansion in inverse powers of area of the horizon cross-section, with coefficients given by derivatives of $F(R)$ wrt $R$ at $R|_{\cal S}=0$. Identifying the leading term of this expansion with the BHAF for $F(R)$ gravity, the rest of the expansion turns into one in inverse powers of the BH entropy. Using the absolute consistency criterion vis-a-vis gravitational wave observations, of corrections to the BHAF, discussed in ref. \cite{pm2025} - \cite{pm2025-2} reviewed briefly in section 2, restrictions ensue on higher ($>1$) order derivatives of $F(R)$ wrt $R$ at $R|_{\cal S}=0$. These constitute parameters of the modified gravity theory. We then point out that the correction terms to the BHAF can be resummed, leading to a concise form of the restrictions on $F(R)$ parameters implied by absolute consistency with GW observations. In section 4, for the sake of comparison, we present a calculation of the first subleading inverse area correction to the BHAF calculated from quantum general relativity, using the It From Bit argument of Wheeler \cite{jaw1991}, modified by ref.\cite{dkm2001} based on tenets of LQG. The coefficient of the inverse power area term in the quantum black hole entropy is seen to obey our absolute consistency criterion.                      

\section{Wald-Jacobson-Kang-Myers Entropy Function}

The Lagrange density, in absence of matter fields. is denoted by $L=L(g_{ab}, R_{abcd})$, where $\nabla$ denotes appropriate spacetime covariant derivatives. A general variation of the classical action yields 
\begin{eqnarray}
\delta L &=& E\cdot \delta \phi + \nabla \cdot \Theta(\delta g_{ab}) \nonumber \\
\Rightarrow \nabla \cdot \Theta &=& \delta L - E_{ab} \delta g_{ab}~. \label{gen}
\end{eqnarray}         
Here, $E$ refers to the equation of motion terms resulting from the general variation of $\phi$, which vanish on-shell.

Consider now infinitesimal diffeomorphisms $x^a \rightarrow x^a + \xi^a(x)$; invariance of the classical action under such transformations imply that \cite{jkm1994}
\begin{eqnarray}
\Theta^a = 2 \frac{\partial L}{\partial R_{abcd}} \nabla_b(\nabla_d \xi_c + \nabla_c \xi_d) + \cdots  \label{diff}
\end{eqnarray}
where the dots imply terms with less than two derivatives on $\xi$. 
The corresponding N\"other current
\begin{eqnarray}
J_N^a &=& \Theta^a - \xi^a L \\ \label{nc}
\Rightarrow \nabla \cdot J_N &=& 0 ~{\rm when}~ E=0 . \label{con}
\end{eqnarray}
For diffeomorphisms, this implies that, once again ignoring less than second order derivative terms of $\xi$ (which is to be identified with the Killing vector field, for stationary isolated black holes), 
\begin{eqnarray}
J_N^a = \nabla_b \left(2 \frac{\partial L}{\partial R_{abcd}} (\nabla_d \xi_c) \right)
\end{eqnarray}
so that the N\"other charge may be written as 
\begin{eqnarray}
Q_N^{ab} = \left(2 \frac{\partial L}{\partial R_{abcd}} \nabla_d \xi_c \right) 
\end{eqnarray}
Identifying $\xi \rightarrow K$, and defining $l_{ab} \equiv \nabla_a K_b$ on the 2-spherical cross-section ${\cal S}$ of the horizon, the Wald entropy function is given by \cite{jkm1994}
\begin{eqnarray}
S_{bh} = \int_{\cal S} d^2a \frac{\partial L}{\partial R_{abcd}} l_{ab} l_{cd} \label{entr}
\end{eqnarray}   

\section{$F(R)$ Gravity}

\subsection{Area Law and corrections}

The motivation for classical modifications of general relativity has been a subject of much debate for decades, and there is no need here to revisit the arguments either in favour or against such modifications. Likewise, the rationale for an $F(R)$ replacement of the Einstein-Hilbert-Lorentz Lagrange density have also been discussed substantively in \cite{bre2004} - \cite{noj2011}. We realize from these important contributions that if a classically modified theory of gravity has black hole solutions like general relativity does, it is likely that the BHAF will receive modifications. For astrophysically macroscopic, spherically symmetric, stationary black hole solutions, with area of the horizon cross section $A_{\cal S} >> A_F$, where, $A_F$ is the analogue of the Planck area for $F(R)$ gravity, and deemed to be a fundamental constant of that theory, it is likely that \cite{pm2025}-\cite{pm2025-2}, the corrections to the area formula analogue of black hole entropy are additive
\begin{eqnarray}
S_{bh} = S_{BH} + s_{bh}(S_{BH}, Q) \label{cor}
\end{eqnarray} 
where, $Q$ represents collectively the parameters of both the theory as well as the black hole solutions under consideration. For astrophysical black holes, one may further stipulate that, since $S_{BH} >> 1$, the corrections admit an expansion in inverse powers of  the horizon area,
\begin{eqnarray}
s_{bh}(S_{BH},Q) = \sum_{n=1} s_n(Q) S_{BH}^{-n} ~. \label{expn} 
\end{eqnarray}  

We now demonstrate the explicit emergence of the BHAF and inverse area corrections for  macroscopic, spherically symmetric black holes of $F(R)$ gravity, using eqn. (\ref{entr}). It is clear that for such theories, we obtain
\begin{eqnarray}
S_{bh} = \int_{\cal S} d^2a F^{(1)}_R(R) l_{ab} l^{ab}~, \label{sfr}
\end{eqnarray} 
where $F_R^{(1)} = dF/dR$. Spherically symmetric static vacuum black holes of $F(R)$ gravity have been explored by \cite{seb2011}, and are characterized by the line element (in Schwarzschild coordinates)
\begin{eqnarray}
ds^2 = B(r) \exp 2\alpha(r) dt^2 - B^{-1}(r) dr^2 - r^2 d\Omega^2 \label{met} 
\end{eqnarray}
The Ricci scalar is given by \cite{seb2011}
\begin{eqnarray}
R(r) &=& 3 B_r \alpha_r + 2B \alpha_r^2 + B_{rr} + 2B \alpha_{rr} \nonumber \\
&+& 4r^{-1} \left[B_r + B \alpha_r \right] + 2r^{-2} (B-1)~. \label{ric}   
\end{eqnarray}
In eqn. (\ref{ric}), a subscript $r$ signifies a first order derivative of that function wrt $r$, while a double subscript signifies a second order derivative. The important point for us is to note that the Ricci scalar is an exclusive function of the radial coordinate $r$. It follows that $R(r_{\cal S}) = R_{\cal S}$, where $r_{\cal S}$ is the radius of the horizon, akin to the Schwarzschild radius for the Schwarzschild black hole in general relativity. Thus, eqn. (\ref{sfr}) can be rewritten as 
\begin{eqnarray}
S_{bh} = F^{(1)}_{R_{\cal S}}(R_{\cal S}) \int_{\cal S} d^2a~ l_{ab} l^{ab} =  F^{(1)}_{R_{\cal S}} (R_{\cal S}) A_{\cal S} ~\label{sfr2}
\end{eqnarray}
where $A_{\cal S}$ is the area of the horizon cross-section. Eqn. (\ref{sfr2}) coincides with eqn. (27) of the review article by Faraoni \cite{far2010}, based on the derivation and interpretations given in ref. \cite{bru2009}. Our eqn (\ref{sfr2}) of course specializes the more general $d$-dimensional result of \cite{bru2009} to four dimensional spherical black hole solutions of general $F(R)$ gravity, so that the inverse effective Newtonian coupling $G_{eff}^{-1}$ is identified with $F^{(1)}_{R_{\cal S}} (R_{\cal S})$. In our derivation, we have used the somewhat easier two codimensional horizon cross-section discussed in ref. \cite{jkm1994}, rather than the foliation of the black hole spacetime by bifurcation spheres of codimension 2, as per the original derivation in ref. \cite{rmw1992}-\cite{iw1994}. The advantage of our rederivation, even though appearing seventeen years following the original, is that it is readily applicable to comparison with astrophysical situations concerning GW data, corresponding to recent observations on BBHCs.    

To this end, observe that we can identify $R_{\cal S}$ in eqn. (\ref{sfr2}) with the {\it Gaussian} curvature of the two dimensional spherical cross-section of the horizon, so that it can be related to the cross-sectional area $A_{\cal S}$
\begin{eqnarray}
R_{\cal S} = \frac{a}{A_{\cal S}} \label{gau}
\end{eqnarray}  
where $a$ is a positive dimensionless number of $O(1)$. Henceforth, we shall absorb this number into a simple redefinition of $F(R)$, exactly as we have absorbed the factor 4 in eqn. (27) of ref. \cite{far2010}. 

In the limit of macroscopic black holes of very large area,  $R_{\cal S}$ is a very small quantity, so that $F_{R_{\cal S}}^{(1)}(R_{\cal S})$ can be Taylor expanded in powers of $R_{\cal S}$ around $R_{\cal S}=0$
\begin{eqnarray}
F^{(1)}_{R_{\cal S}}(R_{\cal S}) &=& \sum_{n=0}\frac{1}{n!} F^{(n+1)}_{R_{\cal S}}(0) R_{\cal S}^n \nonumber \\
&=& F^{(1)}_{R_{\cal S}}(0) + \sum_{n=1}\frac{1}{n!} F^{(n+1)}_{R_{\cal S}}(0) R_{\cal S}^n \label{tay}
\end{eqnarray}
where $  F^{(n)}_{R_{\cal S}}(0)~,~ n=1,2,...$ are the infinitely many parameters of the theory at vanishingly small values of the Ricci curvature.     

Substituting this expansion in eqn (\ref{sfr2}), we obtain
\begin{eqnarray}
S_{bh} = F_{R_{\cal S}}^{(1)}(0) A_{\cal S} + A_{\cal S} \sum_{n=1}\frac{1}{n!} F^{(n+1)}_{R_{\cal S}}(0) R_{\cal S}^n ~. \label{sfr3}
\end{eqnarray}
The first term can be made more akin to the BHAF provided $A_F^{-1} \equiv F^{(1)}_{R_{\cal S}}(0) > 0$; we define $S_{BH} \equiv A_{\cal S}/A_F$. Here, $A_F$ plays the role of the Planck area for $F(R)$ gravity, explicitly analogous to the effective Newtonian {\it constant} of $F(R)$ gravity, when such an identification can be made. This is expected to be small, on general intuitive grounds alone, so that $F^{(1)}_{R_{\cal S}}(0)$ is a large quantity of dimensions of inverse area. One can then define {\it corrections} to the BHAF by splitting off the first term in the expansion (\ref{sfr3}) as the BHAF, and identifying the rest of the expansion with the corrections,  
\begin{eqnarray}
S_{bh} &=& S_{BH} + s_{bh} \nonumber \\
s_{bh} &=& \sum_{n=1} s_{n-1} S_{BH}^{1-n} \nonumber \\
s_{n-1} & \equiv & \frac{1}{n!} [F_{R_{\cal S}}^{(1)}(0)]^{n-1}F_{R_{\cal S}}^{(n+1)}(0) \label{sfr4}
\end{eqnarray}  

\subsection{Generalized Second Law of Thermodynamics and GW Observations}

As discussed in \cite{pm2025}-\cite{pm2025-2}, in the context of binary black hole coalescences (BBHC), the Generalized Second Law of Thermodynamics \cite{bek73} ({\it not the second law of black hole mechanics which is identical to the HAT}) implies, dropping the entropy of the emitted GW, that
\begin{eqnarray}
S_{bhr} & > & S_{bh1} + S_{bh2} {~\rm which} \Rightarrow \nonumber \\
S_{BHr} + s_{bhr} & > & S_{BH1} + S_{BH2} + s_{bh1} + s_{bh2}~. \label{gsl}
\end{eqnarray}
In other words,
\begin{eqnarray}
\Delta S_{BH} &>& - \Delta s_{bh}  \nonumber \\
\Delta S_{BH} & \equiv & S_{BHr} - (S_{BH1} + S_{BH2}) \nonumber \\
\Delta s_{bh} & \equiv & s_{bhr} - (s_{bh1} + s_{bh2}) \label{ineq}
\end{eqnarray}
Careful analyses of GW emission data \cite{teu20} and \cite{bad22}, imply that the data validates Hawking's Area Theorem (HAT) : $ [\Delta S_{BH}]_{obs} > 0$. 

\subsection{Absolute Consistency Criterion}

We now state the {\it absolute consistency} criterion, whereby 
\begin{eqnarray}
\Delta s_{bh} > 0, \label{acc}
\end{eqnarray} 
subject to disclaimers \cite{pm2025}-\cite{pm2025-2}, that $|\Delta s_{bh}| << \Delta S_{BH}$. This is clearly a sufficiency criterion constraining the calculated $\Delta s_{bh}$ by HAT-validating GW data corresponding to BBHC. The alternate {\it relative} consistency criterion : $- \Delta S_{BH} < \Delta s_{bh} < 0$, is already implicit in the fact that $|\Delta s_{bh}| < \Delta S_{BH}$. Thus, the relative consistency criterion is basically without any real predictive power, even though it does not signify any {\it inconsistency} with GW observations. On the other hand, as we shall show, absolute consistency does lead to restrictions on the parameters of $F(R)$ gravity. We may recall that the absolute consistency criterion resembles the HAT in spirit, where the algebraic positive sign of $\Delta A_h$ is far more  important than its magnitude. Likewise, in asymptotically free gauge theories, the sign of the renormalization group $\beta$ function exceeds in importance the magnitude of this function. In fact, as stated in ref. \cite{pm2025-2}, this analogy with the $\beta$ function may be more than just an analogy. 

\subsection{Implications}

Using eqn(\ref{sfr4}), and assuming that the coefficients $s_n$ are identical for similar mass black holes, the absolute consistency criterion implies
\begin{eqnarray}
&& \sum_{n=1}  s_{n-1} \left [ S_{BHr}^{1-n} - S_{BH1}^{1-n} - S_{BH2}^{1-n} \right] > 0 \nonumber \\
\Rightarrow && \sum_{n=1}  s_{n-1} S_{BHr}^{1-n} [ 1 - \left( 1+ \frac{S_{BH1}}{S_{BH2}} \right) ^{n-1} \nonumber \\
&-& \left( 1+ \frac{S_{BH2}}{S_{BH1}} \right) ^{n-1}] > 0 \label{absc}
\end{eqnarray}
We have used the HAT-validating observational result $ \Delta S_{BH} > 0$ in writing down eqn. (\ref{absc}). 

One may object to the above manipulations inside infinite series above, in view of the unknown sign of the coefficient $s_{n-1}$. However, the calculation is reliable for any {\it finite} truncation of the infinite series. In particular, we have for $n=2$, the result
\begin{eqnarray}
s_1 S_{BHr}^{-1} \left[ 1 - \left( 1 + \frac{S_{BH1}}{S_{BH2}} \right) - \left( 1 + \frac{S_{BH2}}{S_{BH1}} \right) \right]  >  0 \label{n2}
\end{eqnarray}      
Observe now that the quantity within the square brackets in the inequality (\ref{n2}) is actually $< -1$, so that one obtains from (\ref{absc}), using also (\ref{sfr4}), 
\begin{eqnarray}
S_{bh} = \frac{A_{\cal S}}{A_F} + F_R^{(2)}(0) + \frac12 F_R^{(3)}(0) A_{\cal S}^{-1} + \cdots, \label{abs2}   
\end{eqnarray}
which leads to, vide eqn (\ref{n2})
\begin{eqnarray}
s_1 = \frac12 F_R^{(3)}(0) F_R^{(1)}(0) < 0 \Rightarrow F_R^{(3)}(0) < 0 . \label{f3}
\end{eqnarray}
Thus if we restrict to the term $n=2$ in inequality (\ref{absc}), as may be valid for remnant black holes in BBHCs with very large horizon area, we obtain the result that $F^{(3)}_{R}(0) < 0$, i.e., for large enough areas of the remnant black hole, we obtain at least an algebraic sign constraint on this parameter of $F(R)$ gravity. This procedure can work for any finite order truncation $n$ of the inequality (\ref{abs2}), which, from a pragmatic standpoint is what may be required practically, depending on the magnitude of the remnant black hole horizon area. Thus, the absolute consistency criterion does indeed impose constraints on the parameters of the theory, as claimed earlier. Note that our focus here has been on the inverse area correction, and hence we have nothing to state about the arbitrary constant $F_R^{(2)}(0)$, given that the entropy functions are always defined upto an added constant.

However, if one wishes, one may sum  the series appearing in (\ref{abs2}); adding and subtracting the $n=0$ term to eqn (\ref{abs2}), and summing the infinite series, we get the inequality 
\begin{eqnarray}
F_{R_{\cal S}}^{(1)} \left( R_{\cal S} + \frac{1}{A_F S_{BHr}} \right)_{R_{\cal S}=0} &<& F_{R_{\cal S}}^{(1)} (0) \label{abs3}
\end{eqnarray}
In other words, we have the inequality relating the parameters of the theory  :
\begin{eqnarray}
F^{(1)}_{R_{\cal S}} (A_{{\cal S}r}^{-1} ) < F_{R_{\cal S}}^{(1)} (0) ~. \label{main}
\end{eqnarray} 
This implies that the first derivative of the function $F$ is a decreasing function of its argument, at least for small positive values of the argument, an inference that stems from our absolute consistency criterion with respect to gravitational wave observations. 

One might seek a clarification as to the limits and scope of our assay above, since the observational data on which (\ref{acc}) is based, deals with general relativistic analyses of {\it spinning} black hole coalescence. On the other hand, the anayses above involves non-spinning spherical black holes of $F(R)$ gravity, and not spinning generalizations thereof. Note however that the entire contentions above involve algebraic sign stipulations of very small, perturbative corrections to the BHAF calculated from the $F(R)$ gravity solution. Had such corrections been such that (\ref{cor}) is no longer valid, then perhaps the spin of the black hole would be an important factor for our consideration. Even so, it is not clear that the algebraic sign stipulations we have made on some $F(R)$ gravity parameters, would be invalidated under those circumstances.   

We now aditionally remark on the {\it scalaron} instability that has been discerned in the original literature on $F(R)$ gravity \cite{sot2010}, and recalled by an anonymous referee in their review. The scalaron field emerges upon a conformal transformation of the $F(R)$ action, transforming it to a {\it scalar-tensor} theory of gravity. It has been shown \cite{sot2010} that the theory can be rendered free of the ghost instability that plagues many higher derivative theories of gravity, by requiring that $F^{(2)}_{R}(0) > 0$. It is clear that HAT-validating GW data together with AC does not impose any restriction on $F^{(2)}_R(0)$ because the corresponding contribution to $S_{bh}$ is independent of $S_{BH}$ or $A_{\cal S}$. This term is therefore an arbitrary constant from our analyses, and in no way therefore contradicts the contentions of ref. \cite{sot2010} on the scalaron instability.

We should however remind the reader that contentions above were based solely on the spherical black hole solution (\ref{met}) for matter-free $F(R)$ gravity. We are not sure what the conformal transformation which maps the $F(R)$ theory to the scalar tensor theory, revealing the scalar degree of freedom, does to the spherical black hole solution (\ref{met}). So we cannot claim from the above analysis presented here that we have discerned any tangible reason for which $F(R)$ gravity is not a viable theory of gravity. Rather, the aim above has been to focus on the GW-observation inspired ACC and see what constraints it might impose on inverse area corrections to the BHAF, and hence indirectly on the parameters of the modified gravity theory. Proving viability of a theory or otherwise would entail a far more detailed investigation beyond the current scope of this letter.    

\section{Subleading corrections to the BHAF : LQG modified It from Bit}

Since inverse area corrections to the BHAF for black hole entropy are not commonly a subject of discussion in the literature, we feel that we may mention another arena where such corrections make their appearance. This is in quantum general relativity. One of the more intuitive approaches towards a quantum general relativity rationale for the BHAF is the It from Bit approach formulated by Wheeler \cite{jaw1991}. In this approach, applicable to spherically symmetric black holes of general relativity, the horizon cross-section 2-sphere of area $A_{\cal S}$ is covered by a spherical lattice of plaquette size $A_{Pl}$, so that the number of plaquettes is $N \simeq A_{\cal S}/A_{Pl}$. On each of these plaquettes is placed a random binary variable, which can be spin $1/2$ variables for example. In this situation, the total number of spin $1/2$ states, i.e., the total number of microstates, on the horizon cross-section is given by 
\begin{eqnarray}
{\cal N} \simeq 2^N = 2^{A_{\cal S}/A_{Pl}}
\end{eqnarray}   

This leads to the Boltzmann entropy 
\begin{eqnarray}
S_{bh} = \log {\cal N} = \frac{A_{\cal S}}{A_{Pl}} ~\log 2 ~\label{jaw}
\end{eqnarray} 
The $\log 2$ coefficient is inevitable if the placing of random variables on the horizon uses binary variables. It can be absorbed into a redefinition of $A_{Pl}$, leading to the BHAF.

This entire scheme fits in nicely with the LQG counting of the microstates for non-rotating isolated black hole horizons \cite{dkm2001}, if {\it all} spins at punctures are assigned the value $1/2$. As is well-known, this contribution counts the number of ways of composing $N$ spin $1/2$ states. However, according to the basic framework of LQG, {\it not all} these states are physical, because they are not singlets under the symmetry group of spin networks namely $SU(2)$. The Gauss law constraint in time gauge (gauge-fixing away the Lorentz boosts) reduces to an $SU(2)$ constraint which only allows singlet states as physical states. Eqn.(\ref{jaw}) therefore represents an {\it overcounting} of the microstates  contributing to $S_{bh}$. 

To obtain the correct counting, consider a large, but even number of plaquettes $N$; one way to obtain a singlet is to count the number of microstates by placing spin $1/2$ components $\pm 1/2$ on half each of the total number of plaquettes. However, this too does not yield the precise number of singlet states, since microstates with total integral spin may also have such a distribution, and these must be excluded. The way to do this is to count the microstates by subtracting off from the $\pm 1/2$ distribution, the states with nonzero total integral spin, all of which have one extra spin $1/2$ component, either positive or negative. Such a counting leads to the formula
\begin{eqnarray}
{\cal N} &=& N! \left[ \frac{1}{[(N/2)!]^2} - \frac{1}{(N/2 +1)! (N/2 -1)!} \right] \nonumber \\
&=& \frac{N!}{(N/2 +1)! (N/2-1)! } \left( \frac{2}{N} \right)
\end{eqnarray}      

We now appeal to the Stirling approximation for the factorials for $N >> 1$,
\begin{eqnarray}
N! \simeq \sqrt{2\pi} \frac{N^{N+1/2}}{e^N} \label{sti}
\end{eqnarray}
This leads to the result
\begin{eqnarray}
{\cal N} & \simeq & \frac{1}{\sqrt{2\pi}} \frac{2^N}{N^{3/2}} \left( 1 + \frac{1}{2N} \right)^{-1} \nonumber \\
\Rightarrow S_{bh} &\simeq& S_{BH} - \frac{3}{2} \log S_{BH} - \frac12 S_{BH}^{-1} + \cdots \label{inva}
\end{eqnarray}
The important point in eqn. (\ref{inva}) is that the coefficient of the subleading inverse horizon area correction is {\it negative} thus in accord with our requirement of absolute consistency for quantum general relativistic black holes (see \cite{pm2024}-\cite{pm2025}). 

\section{Conclusion}

We have considered inverse area corrections to the area formula for black hole entropy, for classical spherically symmetric black holes of $F(R)$ gravity, and for quantum isolated black holes of LQG where they are the subleading corrections following the leading logarithmic corrections discussed in \cite{pm2024}-\cite{pm2025-2}. For $F(R)$ gravity, imposing the criterion of absolute consistency leads to constraints on the parameters of the theory. For the quantum black holes of quantum general relativity formulated as LQG, the subleading correction obeys this criterion. In the former case, this absolute consistency criterion is sufficiently powerful as to constrain the nature of the function $F$ for small positive values of the argument.

Finally, we hark back to the issue raised in the Introduction : how serious are the restrictions on the parameters of $F(R)$ gravity for small arguments. When, with greatly enhanced accuracy of measurement, a particular theory of gravity - general relativity in this case - is seen to emerge as the forerunner, for the correct theory of gravity, it is likely that the observational results will lead to constraints on parameters of a modified gravity approach. If the measurement precision on the validation of HAT from GW data emerging from BBHCs, continues to increase, eventually the modified approaches would be under severe pressure of sustainance. This is the manner in which accurate experimental and observational data aid in the delineation of the correct theoretical framework, across subdisciplines in physics, and there is no reason to believe that gravitational theory should be any different. Our work is an indirect attempt showing how such constraints on the parameters of $F(R)$ gravity may actually emerge on the basis of observational GW data.

\end{document}